# Unruh effect and angular momentum correlation of Rindler particles

Zhigang Bu,[1,*] Liangliang Ji,[1,†] and Baifei Shen[1,2,‡]

[1]*State Key Laboratory of High Field Laser Physics and CAS Center for Excellence in Ultra-intense Laser Science, Shanghai Institute of Optics and Fine Mechanics (SIOM), Chinese Academy of Sciences (CAS), Shanghai 201800, China.*

[2]*Department of Physics, Shanghai Normal University, Shanghai, 200234, China*

**Abstract:** It is well known that the rotational motion does not induce Unruh effect, because the Minkowski vacuum coincides with the vacuum state defined by the pure rotating observers. However, are there Rindler particles carrying orbital angular momentum (OAM), and do they produce observable effect? To answer these questions, we need an accelerated observer having a vortex structure in the transverse dimensions and carrying well-defined OAM. Here we consider the angular momentum characteristics of Unruh effect in the theory of scalar field and electromagnetic field for the first time. We find that the rotation and vortex structure of a uniformly accelerated observer lead to the definite angular momentum correlation between Rindler particles and their counterparts in Minkowski vacuum. When interacting with the Unruh thermal bath, the accelerated vortex observer carrying the OAM of $l$ can absorb and emit Rindler particles with the same OAM, and the particle energy is determined by the angular velocity of the rotation. The absorption rate is not equal to the emission rate in general. Both the absorption and emission processes in the thermal bath are correlated with the emission process of particles with OAM of $\pm l$ in Minkowski vacuum. This effect may promote a potential scheme to detect the Unruh effect. The OAM provides a potential approach to filter out the background noise so that the signal particles could be extracted sufficiently.

e-mail: *zhigang.bu@siom.ac.cn; †jill@siom.ac.cn; ‡ bfshen@mail.shcnc.ac.cn

**I Introduction**

Hawking predicted, in 1974, that black holes could create and emit particles as if they were hot bodies with radiation temperature $T = \hbar\kappa/(2\pi k_B)$ due to the quantum mechanical effects in classical curved space-time [1,2]. This phenomenon is called black hole evaporation. Shortly after Hawking's discovery, Fulling, Davies and Unruh established a closely related effect, the Unruh effect [3-5]: An observer undergoing uniform acceleration experiences the Minkowski vacuum as a thermal bath of particles at temperature $T_{Unruh} = \hbar a/(2\pi k_B)$.

Up to now, the direct observation of Unruh effect is unfeasible because of the requirement for an extreme acceleration undergone by a detector. To manifest a 1K background temperature, the required acceleration would be at least of order $10^{20}$m/s$^2$, which is an enormous challenge with current technology. Still, there have been some proposals for detecting the Unruh effect, including using the sudden ionization of gases or solids on a subpicosecond time scale [6], accelerating electrons in ultrastrong laser fields [7-9], the radiation from the circularly moving charge with constant angular velocity in a uniformly accelerated frame [10], electrons Penning traps [11], atoms in microwave cavities [12], Bose-Einstein condensates [13] or Casimir-Polder force between two accelerating atoms [14]. Schützhold et al. [8,9] noted that the emitted photons from accelerated electrons in inertial frame were always created in correlated pairs, which could be connected to a scattering process of a Rindler photon by the accelerated charge in the uniformly accelerated frame [15]. The similar phenomenon can also happen to scalar particles [16,17]. Besides, Martín-Martínez et al. demonstrated that the Unruh effect may be detected at a lower acceleration using the Berry phase acquired by a pointlike detector moving in space-time [18].

We know that the rotation does not induce Unruh effect although the rotational motion provides normal acceleration. That is because the Minkowski vacuum coincides with the vacuum state defined by the pure rotating observers, who would find the Unruh thermal bath vanishes [19]. Sometimes, the rotation is associated with the vortex carrying well-defined orbital angular momentum (OAM). A question is whether the

OAM would affect the interaction of the observer with the background thermal bath, as well as the radiation properties. In this paper we demonstrate that if an observer has a vortex structure with a spiral azimuthal phase exp(i$l\theta$) in the transverse dimensions, the situation will be completely different, where $l$ is the OAM quantum number, $\theta$ is the azimuth angle. We consider the theory of scalar field and electromagnetic field, and find that although the rotation and vortex does not affect the temperature of the background thermal bath, it introduces the OAM into the interaction, and changes the radiation mode and its spectrum. When the observer interacts with the thermal bath, it induces the radiation emission carrying well-defined OAM in laboratory reference frame (LF). The OAM provides a new and effective degree of freedom to filter out background noise so that the signal could be extracted sufficiently, so it has a potential prospect for the detection of Unruh effect. We shall employ metric signature $(-+++)$, and use units with the Planck constant $\hbar$ and the speed of light $c$ taken to be unity throughout this paper.

**II Scalar field theory**

The Rindler space-time metric is given by $ds^2 = -\exp(2az)dt^2 + dr^2 + r^2 d\theta^2 + \exp(2az)dz^2$, where $a$ is the uniform proper acceleration and treated as a positive constant. In this paper, we concentrate on the OAM characteristics associated with the vortex, so it is more convenient to transform to the cylindrical coordinate, where the Rindler metric takes the form

$$ds^2 = -e^{2az}dt^2 + dr^2 + r^2 d\theta^2 + e^{2az}dz^2. \tag{2.1}$$

The coordinates ($t$, $r$, $\theta$, $z$) cover two regions where the killing vector $Z\partial_T + T\partial_Z$ is timelike: the right Rindler wedge with |$T$|<$Z$, and the left Rindler wedge with |$T$|<-$Z$, where the capitals denote the coordinates in Minkowski reference frame (MF). For the right Rindler wedge, metric (2.1) leads to the coordinate transformation between Rindler reference frame (RF) and MF

$$T = a^{-1}e^{az}\sinh(at), \quad R = r, \quad \Theta = \theta, \quad Z = a^{-1}e^{az}\cosh(at). \tag{2.2}$$

## 2.1 The angular momentum eigen mode and field quantization in RF

The Lagrangian density of a scalar field is given by

$$\mathcal{L}(\Psi, \partial_\mu \Psi) = \frac{\sqrt{-g}}{2}\left[g^{\mu\nu}(\partial_\mu \Psi)(\partial_\nu \Psi) + M^2 \Psi^2\right], \tag{2.3}$$

where $M$ is the mass of the scalar particle, $g$ is the metric determinant. Under the variational principle, we obtain the Klein-Gordon equation

$$(\nabla_\mu \nabla^\mu - M^2)\Psi(x^\nu) = 0. \tag{2.4}$$

In the right Rindler wedge, the solution of Eq. (2.4) is

$$\psi^R_{m,\kappa,\omega}(x^\mu) = \sqrt{\frac{1}{2\pi^3 a}\sinh\left(\frac{\pi\omega}{a}\right)}\, e^{im\theta - i\omega t} J_m(\kappa r) K_{i\omega/a}\left(a^{-1}(\kappa^2 + M^2)^{1/2} e^{az}\right), \tag{2.5}$$

where $m \in \mathbb{Z}$ and $\kappa, \omega > 0$, $J_m(s)$ is the Bessel function of the first kind, $K_\nu(s)$ is the modified Bessel function. Using the definition of the inner product between two modes: $(\psi_{m',\kappa',\omega'}, \psi_{m,\kappa,\omega}) = -i\int \sqrt{-g}\, g^{0\nu}(\psi^*_{m',\kappa',\omega'}\partial_\nu \psi_{m,\kappa,\omega} - \psi_{m,\kappa,\omega}\partial_\nu \psi^*_{m',\kappa',\omega'})dr d\theta dz$, mode (2.5) satisfies the orthonormality and normalization conditions

$$(\psi^R_{m',\kappa',\omega'}, \psi^R_{m,\kappa,\omega}) = -(\psi^{R*}_{m',\kappa',\omega'}, \psi^{R*}_{m,\kappa,\omega}) = \frac{1}{\kappa}\delta_{mm'}\delta(\kappa - \kappa')\delta(\omega - \omega'), \tag{2.6a}$$

$$(\psi^{R*}_{m',\kappa',\omega'}, \psi^R_{m,\kappa,\omega}) = (\psi^R_{m',\kappa',\omega'}, \psi^{R*}_{m,\kappa,\omega}) = 0. \tag{2.6b}$$

In right Rindler wedge, a scalar field $\Psi$ can be expanded as

$$\Psi = \sum_m \int d\omega d\kappa\, \kappa \left(a^R_{m,\kappa,\omega}\psi^R_{m,\kappa,\omega} + a^{R\dagger}_{m,\kappa,\omega}\psi^{R*}_{m,\kappa,\omega}\right), \tag{2.7}$$

where $a^R_{m,\kappa,\omega} = (\psi^R_{m,\kappa,\omega}, \Psi)$ and $a^{R\dagger}_{m,\kappa,\omega} = (\Psi, \psi^R_{m,\kappa,\omega})$ are annihilation and creation operators and satisfy the commutation relations,

$$\left[a^R_{m',\kappa',\omega'},\, a^{R\dagger}_{m,\kappa,\omega}\right] = \frac{1}{\kappa}\delta_{mm'}\delta(\kappa - \kappa')\delta(\omega - \omega'), \tag{2.8a}$$

$$\left[a^R_{m',\kappa',\omega'},\, a^R_{m,\kappa,\omega}\right] = \left[a^{R\dagger}_{m',\kappa',\omega'},\, a^{R\dagger}_{m,\kappa,\omega}\right] = 0. \tag{2.8b}$$

The similar procedure can be performed to derived the eigen mode $\psi^L_{m,\kappa,\omega}$ and field quantization in the left Rindler wedge, with the annihilation and creation operators $a^L_{m,\kappa,\omega}$ and $a^{L\dagger}_{m,\kappa,\omega}$. The Rindler vacuum state, $|0_R\rangle$, is defined by requiring that $a^R_{m,\kappa,\omega}|0_R\rangle = a^L_{m,\kappa,\omega}|0_R\rangle = 0$ for all $m$, $\kappa$ and $\omega$.

## 2.2 Unruh effect

In Minkowski space-time, the Klein-Gordon equation is $\left[-\partial_T^2 + (1/R)\partial_R(R\partial_R) + (1/R^2)\partial_\Theta^2 + \partial_Z^2 - M^2\right]\varphi(T,R,\Theta,Z) = 0$. The solution,

$$\varphi_{m_0,\kappa_0,k_Z,\omega_0}(T,R,\Theta,Z) = \frac{1}{2\pi}\sqrt{\frac{1}{2\omega_0}}e^{im_0\Theta + ik_Z Z - i\omega_0 T} J_{m_0}(\kappa_0 R), \qquad (2.9)$$

satisfies the orthonormality and normalization conditions. Let us define the modes by using the linear combinations of the Rindler modes $\psi_{m,\kappa,\omega}^R$ and $\psi_{m,\kappa,\omega}^L$:

$$\phi_{m,\kappa,\omega} = \frac{\psi_{m,\kappa,\omega}^R + \exp\left(-\dfrac{\pi\omega}{a} + im\pi\right)\psi_{-m,\kappa,\omega}^{L*}}{\left[1 - \exp\left(-\dfrac{2\pi\omega}{a}\right)\right]^{1/2}}, \qquad (2.10a)$$

$$\bar{\phi}_{m,\kappa,\omega} = \frac{\psi_{m,\kappa,\omega}^L + \exp\left(-\dfrac{\pi\omega}{a} + im\pi\right)\psi_{-m,\kappa,\omega}^{R*}}{\left[1 - \exp\left(-\dfrac{2\pi\omega}{a}\right)\right]^{1/2}}. \qquad (2.10b)$$

It can be proven that $\phi_{m,\kappa,\omega}$ and $\bar{\phi}_{m,\kappa,\omega}$ are superpositions of purely positive-frequency modes, Eq. (2.9), in Minkowski space-time, and satisfies the orthonormality and normalization conditions. From Eq. (2.10) we obtain

$$\psi_{m,\kappa,\omega}^R = \frac{\phi_{m,\kappa,\omega} - \exp\left(-\dfrac{\pi c\omega}{a} + im\pi\right)\bar{\phi}_{-m,\kappa,\omega}^{*}}{\left[1 - \exp\left(-\dfrac{2\pi c\omega}{a}\right)\right]^{1/2}}, \qquad (2.11a)$$

$$\psi_{m,\kappa,\omega}^L = \frac{\bar{\phi}_{m,\kappa,\omega} - \exp\left(-\dfrac{\pi c\omega}{a} + im\pi\right)\phi_{-m,\kappa,\omega}^{*}}{\left[1 - \exp\left(-\dfrac{2\pi c\omega}{a}\right)\right]^{1/2}}. \qquad (2.12b)$$

Now a scalar field $\Psi$ can be expanded in Minkowski positive-frequency modes $\phi_{m,\kappa,\omega}$ as

$$\Psi = \sum_m \int d\omega \kappa d\kappa \left(b_{m,\kappa,\omega}\phi_{m,\kappa,\omega} + b_{m,\kappa,\omega}^\dagger \phi_{m,\kappa,\omega}^{*}\right), \qquad (2.13)$$

with the annihilation and creation operators, $b_{m,\kappa,\omega}$ and $b_{m,\kappa,\omega}^\dagger$, satisfying

$$\left[b_{m',\kappa',\omega'}, b^\dagger_{m,\kappa,\omega}\right] = (1/\kappa)\delta_{mm'}\delta(\kappa-\kappa')\delta(\omega-\omega') \quad \text{and}$$

$\left[b_{m',\kappa',\omega'}, b_{m,\kappa,\omega}\right] = \left[b^\dagger_{m',\kappa',\omega'}, b^\dagger_{m,\kappa,\omega}\right] = 0$. The relationship of annihilation and creation operators between the Rindler modes and Minkowski modes is derived as,

$$a^R_{m,\kappa,\omega} = \frac{b_{m,\kappa,\omega} + \exp\left(-\frac{\pi\omega}{a} + im\pi\right) b^\dagger_{-m,\kappa,-\omega}}{\left[1 - \exp\left(-\frac{2\pi\omega}{a}\right)\right]^{1/2}}, \quad a^{R\dagger}_{m,\kappa,\omega} = \frac{b^\dagger_{m,\kappa,\omega} + \exp\left(-\frac{\pi\omega}{a} - im\pi\right) b_{-m,\kappa,-\omega}}{\left[1 - \exp\left(-\frac{2\pi\omega}{a}\right)\right]^{1/2}},$$

(2.14a)

$$a^L_{m,\kappa,\omega} = \frac{b_{m,\kappa,-\omega} + \exp\left(-\frac{\pi\omega}{a} + im\pi\right) b^\dagger_{-m,\kappa,\omega}}{\left[1 - \exp\left(-\frac{2\pi\omega}{a}\right)\right]^{1/2}}, \quad a^{L\dagger}_{m,\kappa,\omega} = \frac{b^\dagger_{m,\kappa,-\omega} + \exp\left(-\frac{\pi\omega}{a} - im\pi\right) b_{-m,\kappa,\omega}}{\left[1 - \exp\left(-\frac{2\pi\omega}{a}\right)\right]^{1/2}}.$$

(2.14b)

Using Eqs. (2.14), the expectation value of the Rindler particle number operator, $\hat{N}^R_{m,\kappa,\omega} = a^{R\dagger}_{m,\kappa,\omega} a^R_{m,\kappa,\omega}$, in Minkowski vacuum is calculated as

$$\left\langle 0_M \left| \hat{N}^R_{m,\kappa,\omega} \right| 0_M \right\rangle = \sum_{m'} \int d\omega' d\kappa' \kappa' \left\langle 0_M \left| a^{R\dagger}_{m',\kappa',\omega'} a^R_{m,\kappa,\omega} \right| 0_M \right\rangle = \frac{1}{\exp(2\pi\omega/a) - 1}. \quad (2.15)$$

For left Rindler wedge, we can obtain the same result. This indicates that the Minkowski vacuum state is a thermal equilibrium state when it is probed in RF with the temperature $T = a/(2\pi)$. Therefore, an observer undergoing uniform proper acceleration experiences the Minkowski vacuum as a thermal bath. The background temperature is only dependent on the proper acceleration, and not affected by the OAM. However, the OAM can induce the physical effects when the observer interacts with the background thermal bath, as shown in next section.

## 2.3 Physical effects in Minkowski reference frame (MF)
### 2.3.1 Absorption and emission rates from a vortex observer in thermal bath

To demonstrate the possible observable effect in laboratory reference frame, we consider an observer moving along a hyperbolic world line with a uniform proper acceleration $a$. The interaction Hamiltonian of the observer with the thermal bath is expressed as $H_{int} = \varepsilon \int h(x)\Psi(x)\sqrt{-g(x)}d^3x$, where $h(x)$ is a scalar function determining the characteristic of the observer, $g(x)$ is the metric determinant, and $\varepsilon$ is the interaction coefficient. In this paper, we will focus on the angular momentum effect on Unruh effect, so we suppose $h(x)$ has a rotating vortex structure in the transversal dimensions and is expressed by $h(x) = (1/2)(r/w)^{|l|}\exp(-r^2/w^2)\delta(z)(1+\cos(l\theta - \Omega t))$, where $l$ is the OAM quantum number, $\Omega>0$ determines the constant angular velocity of the vortex in RF, $w$ is the waist radius. We find the first term of $h(x)$ is static and independent of $\theta$, so it behaves as a conventional observer and does not interact with the particles carrying OAM in the thermal bath. We are only interested in the vortex term taking the form of

$$h_{vor}(x) = \frac{1}{2}\left(\frac{r}{w}\right)^{|l|}\exp\left(-\frac{r^2}{w^2}\right)\delta(z)\cos(l\theta - \Omega t), \qquad (2.16)$$

which will produce the angular momentum effects.

The S-operator of the interaction is $\hat{S}(t,t_0) = T\exp\left[-i\varepsilon\int_{t_0}^{t} h(x')\Psi(x')\sqrt{-g(x')}d^3x'dt'\right]$. In the first order approximation, the amplitude for absorbing a particle with mode $|m,\kappa,\omega\rangle_R$ from the thermal bath by the observer is given by $S_{absorb}^{R}(m,\kappa,\omega) = -i\varepsilon\sqrt{N_{m,\kappa,\omega}}\int d^4x\sqrt{-g(x)}h_{vor}(x)\psi_{m,\kappa,\omega}^{R}(x)$, where $N_{m,\kappa,\omega}$ denotes the particle number with mode $|m,\kappa,\omega\rangle_R$ in the Unruh thermal bath determined by Eq. (2.15). Using Eqs. (2.5) and (2.16) we obtain the absorption amplitude

$$S_{absorb}^{R}(m,\kappa,\omega) = -\frac{i\pi^2\varepsilon\kappa^{|l|}w^{|l|+2}}{2^{|l|+1}}\sqrt{\frac{N_{m,\kappa,\omega}}{2\pi^3 a}\sinh\left(\frac{\pi\omega}{a}\right)}\exp\left(-\frac{\kappa^2 w^2}{4}\right) \\ \times K_{i\omega/a}\left(\frac{1}{a}(\kappa^2+M^2)^{1/2}\right)\left(\frac{l}{|l|}\right)^{|l|}\delta_{ml}\delta(\omega-\Omega) \qquad (2.17)$$

The amplitude for emitting a particle with mode $|m,\kappa,\omega\rangle_R$ to the thermal bath by the observer is calculated as $S_{emit}^{R}(m,\kappa,\omega) = -i\varepsilon\sqrt{N_{m,\kappa,\omega}+1}\int d^4x\sqrt{-g(x)}h_{vor}(x)\psi_{m,\kappa,\omega}^{R*}(x)$,

with the result,

$$S_{emit}^{R}(m,\kappa,\omega) = -\frac{i\pi^{2}\varepsilon\kappa^{|l|}w^{|l|+2}}{2^{|l|+1}}\sqrt{\frac{(N_{m,\kappa,\omega}+1)}{2\pi^{3}a}\sinh\left(\frac{\pi\omega}{a}\right)}\exp\left(-\frac{\kappa^{2}w^{2}}{4}\right) \\ \times K_{i\omega/a}\left(\frac{1}{a}(\kappa^{2}+M^{2})^{1/2}\right)\left(\frac{l}{|l|}\right)^{|l|}\delta_{ml}\delta(\omega-\Omega)$$

(2.18)

From Eqs. (2.17) and (2.18) we find that if the observer itself carries a well-defined OAM of $l$, it can absorb and emit Rindler particles with the same OAM of $l$ when it interacts with the background thermal bath, and the energy of the absorbed/emitted Rindler particles is determined by the rotational energy of the observer, $\omega=\Omega$. The differential absorption rate per unit time can be calculated from

$$P_{absorb}^{R}=\frac{1}{T}\sum_{m}\int d\omega d\kappa\kappa\left|S_{absorb}^{R}(m,\kappa,\omega)\right|^{2},$$ with the result

$$\frac{dP_{absorb}^{R}}{d\kappa}=\frac{\varepsilon^{2}w^{2|l|+4}}{2^{2|l|+5}a}\exp\left(-\frac{\pi\Omega}{a}\right)\kappa^{2|l|+1}\exp\left(-\frac{\kappa^{2}w^{2}}{2}\right)K_{i\Omega/a}^{2}\left(\frac{1}{a}(\kappa^{2}+M^{2})^{1/2}\right),$$ (2.19)

and the emission rate per unit time is given by

$$\frac{dP_{emit}^{R}}{d\kappa}=\frac{\varepsilon^{2}w^{2|l|+4}}{2^{2|l|+5}a}\exp\left(\frac{\pi\Omega}{a}\right)\kappa^{2|l|+1}\exp\left(-\frac{\kappa^{2}w^{2}}{2}\right)K_{i\Omega/a}^{2}\left(\frac{1}{a}(\kappa^{2}+M^{2})^{1/2}\right).$$ (2.20)

We find the absorption rate is not equal to the emission rate when the observer carries OAM. Only when the angular velocity $\Omega\to 0$, the absorption rate equals to the emission rate.

### 2.3.2 Emission rate of the observer in Minkowski vacuum

In Minkowski space-time, the scalar function $h_{vor}(x)$ becomes

$$h_{vor}(X)=\frac{1}{2}\left(\frac{R}{w}\right)^{|l|}\exp\left(-\frac{R^{2}}{w^{2}}\right)\delta(z)\cos(l\Theta-\Omega t),$$ (2.21)

with $z=(1/(2a))\ln\left[a^{2}(Z^{2}-T^{2})\right]$ and $t=(1/a)\mathrm{arctanh}(T/Z)$. The amplitude for emitting a particle with mode $|m,\kappa,k_{z},\omega\rangle_{M}$ to the Minkowski vacuum by the accelerated observer is given by $S_{emit}^{M}(m,\kappa,k_{z},\omega)=-i\varepsilon\int d^{4}X\sqrt{-g(X)}h_{vor}(X)\varphi_{m,\kappa,k_{z},\omega}^{*}(X)$. Using the mode (2.9), the emission amplitude in Minkowski vacuum is derived,

$$S_{emit}^{M}(m,\kappa,k_z,\omega) = -\frac{i\varepsilon\kappa^{|l|}w^{|l|+2}}{2^{|l|+2}a\sqrt{2\omega}}\exp\left(-\frac{\kappa^2 w^2}{4}\right)K_{i\Omega/a}\left(\frac{1}{a}(\kappa^2+M^2)^{1/2}\right)$$

$$\times\left[\delta_{ml}\left(\frac{l}{|l|}\right)^{|l|}\left(\frac{\omega+k_z}{\omega-k_z}\right)^{-i\Omega/(2a)}\exp\left(\frac{\pi\Omega}{2a}\right)+\delta_{m,-l}\left(-\frac{l}{|l|}\right)^{|l|}\left(\frac{\omega+k_z}{\omega-k_z}\right)^{i\Omega/(2a)}\exp\left(-\frac{\pi\Omega}{2a}\right)\right].$$

(2.22)

Eq. (2.22) contains two terms with Kronecker deltas, $\delta_{ml}$ and $\delta_{m,-l}$, respectively, which indicates the emitted particles carry the well-defined OAM of $m=\pm l$. Then the differential rate for emitting a particle from the vortex observer to the Minkowski vacuum per unit time can be calculated, with the result

$$\frac{dP_{emit}^{M}}{d\kappa} = \frac{\varepsilon^2 w^{2|l|+4}}{2^{2|l|+4}a}\cosh\left(\frac{\pi\Omega}{a}\right)\kappa^{2|l|+1}\exp\left(-\frac{\kappa^2 w^2}{2}\right)K_{i\Omega/a}^{2}\left(\frac{1}{a}(\kappa^2+M^2)^{1/2}\right). \quad (2.23)$$

Now we find $P_{emit}^{M}(m=\pm l) = P_{absorb}^{R}(m=l) + P_{emit}^{R}(m=l)$, which means if the observer has a rotating vortex structure and carries a OAM of $l$, it would absorb or emit the Rindler particles with energy of $\omega=\Omega$ and OAM of $m=l$ in the thermal bath, and the sum of the absorption and emission rates is equal to the total emission rate of particles with OAM of $m=\pm l$ in Minkowski vacuum. However, the absorption rate is not identical to the emission rate in background thermal bath. From the two terms in Eq. (2.22), we also notice that the emission rate of Rindler particles with OAM of $l$ is equal to the emission rate of particles carrying the same OAM in Minkowski vacuum, $P_{emit}^{M}(m=l) = P_{emit}^{R}(m=l)$, but the absorption rate of Rindler particles is equal to the emission rate of particles with opposite OAM in Minkowski vacuum, $P_{emit}^{M}(m=-l) = P_{absorb}^{R}(m=l)$.

**III Electromagnetic field theory**

**3.1 The angular momentum eigen mode and field quantization in RF**

The Lagrangian density of an electromagnetic field is

$$\mathcal{L} = -\frac{1}{4}\sqrt{-g}F_{\mu\nu}F^{\mu\nu} - \frac{1}{2\varsigma}\sqrt{-g}(\nabla_\nu A^\nu)^2. \quad (3.1)$$

In the Feynman gauge, $\varsigma = 1$, the electromagnetic field equation can be obtained from the variational principle: $\left(A_{\mu;\nu}\right)^{\nu}_{;} = \nabla_{\nu}\nabla^{\nu}A_{\mu} = 0$. We impose the Lorentz gauge, $\nabla^{\mu}A_{\mu} = 0$, for the physical modes, under which the field equation has two solutions describing two physical modes [10,20]:

$$A^{(1)}_{\mu;m,\kappa,\omega}(x) = \left(\frac{\partial f_{m,\kappa,\omega}(x)}{\partial z},\ 0,\ 0,\ \frac{\partial f_{m,\kappa,\omega}(x)}{\partial t}\right), \tag{3.2a}$$

$$A^{(2)}_{\mu;m,\kappa,\omega}(x) = \left(0,\ -\frac{1}{r}\frac{\partial f_{m,\kappa,\omega}(x)}{\partial \theta},\ r\frac{\partial f_{m,\kappa,\omega}(x)}{\partial r},\ 0\right), \tag{3.2b}$$

where the scalar function $f_{m,\kappa,\omega}(x)$ satisfies

$$e^{-2az}\left(\frac{\partial^2 f_{m,\kappa,\omega}}{\partial t^2} - \frac{\partial^2 f_{m,\kappa,\omega}}{\partial z^2}\right) - \frac{1}{r}\frac{\partial}{\partial r}\left(r\frac{\partial f_{m,\kappa,\omega}}{\partial r}\right) - \frac{1}{r^2}\frac{\partial^2 f_{m,\kappa,\omega}}{\partial \theta^2} = 0. \tag{3.3}$$

The solution of Eq. (3.3) is

$$f_{m,\kappa,\omega}(t,r,\theta,z) = F_{m,\kappa,\omega} e^{im\theta - i\omega t} J_m(\kappa r) K_{i\omega/a}\left(\frac{\kappa}{a}e^{az}\right), \tag{3.4a}$$

with the normalization constant $F_{m,\kappa,\omega} = \sqrt{1/(2\pi^3 a\kappa^2)\sinh(\pi\omega/a)}$. We also need another two modes to form a complete basis for the electromagnetic field. There is a pure gauge mode satisfying the Lorentz gauge: $A^{(G)}_{\mu}(x) = \left(\partial_t f_{m,\kappa,\omega},\ \partial_r f_{m,\kappa,\omega},\ \partial_\theta f_{m,\kappa,\omega},\ \partial_z f_{m,\kappa,\omega}\right)$. We also have a mode only satisfying the field equation: $A^{(W)}_{\mu}(x) = \left(0,\ \partial_r f_{m,\kappa,\omega},\ \partial_\theta f_{m,\kappa,\omega},\ 0\right)$. Then we can construct another two pure gauge and nonphysical modes as follows,

$$A^{(0)}_{\mu}(x) = A^{(G)}_{\mu}(x) - A^{(W)}_{\mu}(x) = \left(\frac{\partial f_{m,\kappa,\omega}(x)}{\partial t},\ 0,\ 0,\ \frac{\partial f_{m,\kappa,\omega}(x)}{\partial z}\right), \tag{3.2c}$$

$$A^{(3)}_{\mu}(x) = A^{(W)}_{\mu}(x) = \left(0,\ \frac{\partial f_{m,\kappa,\omega}(x)}{\partial r},\ \frac{\partial f_{m,\kappa,\omega}(x)}{\partial \theta},\ 0\right). \tag{3.2d}$$

Using the definition of the inner product between two modes [15], we find that the solutions (3.2) satisfy the orthonormality and normalization conditions:

$$\left(A^{(\sigma')}_{m',\kappa',\omega'},\ A^{(\sigma)}_{m,\kappa,\omega}\right) = -\left(A^{(\sigma')*}_{m',\kappa',\omega'},\ A^{(\sigma)*}_{m,\kappa,\omega}\right) = \frac{\eta_{\sigma\sigma'}}{\kappa}\delta_{mm'}\delta(\kappa-\kappa')\delta(\omega-\omega'), \tag{3.5a}$$

$$\left(A^{(\sigma')}_{m',\kappa',\omega'}\ ,\ A^{(\sigma)*}_{m,\kappa,\omega}\right)=\left(A^{(\sigma')*}_{m',\kappa',\omega'}\ ,\ A^{(\sigma)}_{m,\kappa,\omega}\right)=0 \tag{3.5b}$$

where $\sigma=1,2$ represents the two physical modes, $\sigma=0,3$ represents the two nonphysical modes, and $\eta_{\sigma\sigma'}=\mathrm{diag}(-1,\ 1,\ 1,\ 1)$. Up to now, our discussion is performed in the right Rindler wedge. For the left Rindler wedge, the solutions have the same expressions as modes (3.2), but the scalar function $f_{m,\kappa,\omega}(x)$ should be replaced with $\bar{f}_{m,\kappa,\omega}(\bar{x})$,

$$\bar{f}\left(\bar{t},\bar{r},\bar{\theta},\bar{z}\right)=F_{m,\kappa,\omega}e^{im\bar{\theta}+i\omega\bar{t}}J_m\left(\kappa\bar{r}\right)K_{i\omega/a}\left(\frac{\kappa}{a}e^{a\bar{z}}\right). \tag{3.4b}$$

In right Rindler wedge, an electromagnetic field $\mathscr{A}_\mu$ can be expanded in the basis of $A^{(\sigma)R}_{\mu;m,\kappa,\omega}$,

$$\mathscr{A}_\mu=\sum_{\sigma=0}^{3}\sum_{m=-\infty}^{\infty}\int d\omega d\kappa \kappa\left(a^{R(\sigma)}_{m,\kappa,\omega}A^{R(\sigma)}_{\mu;m,\kappa,\omega}+a^{R(\sigma)\dagger}_{m,\kappa,\omega}A^{R(\sigma)*}_{\mu;m,\kappa,\omega}\right), \tag{3.6}$$

the corresponding annihilation and creation operators are determined by $a^{R(\sigma)}_{m,\kappa,\omega}=\zeta_\sigma\left(A^{R(\sigma)}_{m,\kappa,\omega}\ ,\ \mathscr{A}\right)$ and $a^{R(\sigma)\dagger}_{m,\kappa,\omega}=\zeta_\sigma\left(\mathscr{A}\ ,\ A^{R(\sigma)}_{m,\kappa,\omega}\right)$, and satisfy the commutation relations,

$$\left[a^{R(\sigma')}_{m',\kappa',\omega'}\ ,\ a^{R(\sigma)\dagger}_{m,\kappa,\omega}\right]=\frac{\eta_{\sigma\sigma'}}{\kappa}\delta_{mm'}\delta(\kappa-\kappa')\delta(\omega-\omega'), \tag{3.7a}$$

$$\left[a^{R(\sigma')}_{m',\kappa',\omega'}\ ,\ a^{R(\sigma)}_{m,\kappa,\omega}\right]=\left[a^{R(\sigma')\dagger}_{m',\kappa',\omega'}\ ,\ a^{R(\sigma)\dagger}_{m,\kappa,\omega}\right]=0, \tag{3.7b}$$

with the coefficient $\zeta_\sigma=+1$ for $\sigma\neq 0$, and $\zeta_\sigma=-1$ for $\sigma=0$. The similar procedure can be performed to quantize the field in the left Rindler wedge, with the annihilation and creation operators $a^{L(\sigma)}_{m,\kappa,\omega}$ and $a^{L(\sigma)\dagger}_{m,\kappa,\omega}$. The Rindler vacuum state, $|0_R\rangle$, is defined by requiring that $a^{R(\sigma)}_{m,\kappa,\omega}|0_R\rangle=a^{L(\sigma)}_{m,\kappa,\omega}|0_R\rangle=0$ for all $\sigma$, $m$, $\kappa$ and $\omega$. Assuming that $\left|N^{R(\sigma)}_{m,\kappa,\omega}\right\rangle$ is the eigen state of the particle number operator, $\hat{N}^{R(\sigma)}_{m,\kappa,\omega}=\zeta_\sigma a^{R(\sigma)\dagger}_{m,\kappa,\omega}a^{R(\sigma)}_{m,\kappa,\omega}$, with the eigen value $N^{R(\sigma)}_{m,\kappa,\omega}$, $\hat{N}^{R(\sigma)}_{m,\kappa,\omega}\left|N^{R(\sigma)}_{m,\kappa,\omega}\right\rangle=N^{R(\sigma)}_{m,\kappa,\omega}\left|N^{R(\sigma)}_{m,\kappa,\omega}\right\rangle$. When an annihilation or creation operator acts on $\left|N^{R(\sigma)}_{m,\kappa,\omega}\right\rangle$, we have

$$a_{m,\kappa,\omega}^{R(\sigma)}\left|N_{m,\kappa,\omega}^{R(\sigma)}\right\rangle = \zeta_\sigma \sqrt{N_{m,\kappa,\omega}^{R(\sigma)}}\left|N_{m,\kappa,\omega}^{R(\sigma)}-1\right\rangle \text{ or } a_{m,\kappa,\omega}^{R(\sigma)\dagger}\left|N_{m,\kappa,\omega}^{R(\sigma)}\right\rangle = \sqrt{N_{m,\kappa,\omega}^{R(\sigma)}+1}\left|N_{m,\kappa,\omega}^{R(\sigma)}+1\right\rangle.$$

## 3.2 Unruh effect of photon

Noting that the right Rindler modes, $A_{m,\kappa,\omega}^{R(\sigma)}$, are zero in the left Rindler wedge and left Rindler modes, $A_{m,\kappa,\omega}^{L(\sigma)}$, are zero in the right Rindler wedge, one can construct a set of Minkowski modes as follows:

$$\Phi_{\mu;m,\kappa,\omega}^{(\sigma)} = \frac{A_{\mu;m,\kappa,\omega}^{R(\sigma)} + \exp\left(-\frac{\pi\omega}{a} + im\pi\right)A_{\mu;-m,\kappa,\omega}^{L(\sigma)*}}{\left[1-\exp\left(-\frac{2\pi\omega}{a}\right)\right]^{1/2}}, \quad (3.8a)$$

$$\bar{\Phi}_{\mu;m,\kappa,\omega}^{(\sigma)} = \frac{A_{\mu;m,\kappa,\omega}^{L(\sigma)} + \exp\left(-\frac{\pi\omega}{a} + im\pi\right)A_{\mu;-m,\kappa,\omega}^{R(\sigma)*}}{\left[1-\exp\left(-\frac{2\pi\omega}{a}\right)\right]^{1/2}}. \quad (3.8b)$$

According to the vectorial transformation rules, one can derive the corresponding electromagnetic modes in MF from the Rindler modes under the transformation: $A_\mu^{M(\sigma)}(X) = (\partial x^\nu/\partial X^\mu)A_\nu^{(\sigma)}(\Lambda^{-1}x)$. Thus, we define a set of modes $\Phi'^{(\sigma)}_{\mu;m,\kappa,\omega} = \left[1-\exp(-2\pi\omega/a)\right]^{1/2}\Phi^{(\sigma)}_{\mu;m,\kappa,\omega}$, they take the forms

$$\Phi'^{(0)}_{\mu;m,\kappa,\omega} = \begin{pmatrix}\partial_T \tilde{f}_{m,\kappa,\omega}(X)\\0\\0\\\partial_Z \tilde{f}_{m,\kappa,\omega}(X)\end{pmatrix}\varepsilon(-U) - i\exp\left(-\frac{\pi\omega}{a}\right)\begin{pmatrix}\partial_T \tilde{\bar{f}}_{m,\kappa,-\omega}(X)\\0\\0\\\partial_Z \tilde{\bar{f}}_{m,\kappa,-\omega}(X)\end{pmatrix}\varepsilon(U), \quad (3.9a)$$

$$\Phi'^{(1)}_{\mu;m,\kappa,\omega} = \begin{pmatrix}\partial_Z \tilde{f}_{m,\kappa,\omega}(X)\\0\\0\\\partial_T \tilde{f}_{m,\kappa,\omega}(X)\end{pmatrix}\varepsilon(-U) - i\exp\left(-\frac{\pi\omega}{a}\right)\begin{pmatrix}\partial_Z \tilde{\bar{f}}_{m,\kappa,-\omega}(X)\\0\\0\\\partial_T \tilde{\bar{f}}_{m,\kappa,-\omega}(X)\end{pmatrix}\varepsilon(U), \quad (3.9b)$$

$$\Phi'^{(2)}_{\mu;m,\kappa,\omega} = \begin{pmatrix}0\\-R^{-1}\partial_\Theta \tilde{f}_{m,\kappa,\omega}(X)\\R\partial_R \tilde{f}_{m,\kappa,\omega}(X)\\0\end{pmatrix}\varepsilon(-U) - i\exp\left(-\frac{\pi\omega}{a}\right)\begin{pmatrix}0\\-R^{-1}\partial_\Theta \tilde{\bar{f}}_{m,\kappa,-\omega}(X)\\R\partial_R \tilde{\bar{f}}_{m,\kappa,-\omega}(X)\\0\end{pmatrix}\varepsilon(U), \quad (3.9c)$$

$$\Phi'^{(3)}_{\mu;m,\kappa,\omega} = \begin{pmatrix} 0 \\ \partial_R \tilde{f}_{m,\kappa,\omega}(X) \\ \partial_\Theta \tilde{f}_{m,\kappa,\omega}(X) \\ 0 \end{pmatrix} \varepsilon(-U) - i\exp\left(-\frac{\pi\omega}{a}\right) \begin{pmatrix} 0 \\ \partial_R \tilde{\tilde{f}}_{m,\kappa,-\omega}(X) \\ \partial_\Theta \tilde{\tilde{f}}_{m,\kappa,-\omega}(X) \\ 0 \end{pmatrix} \varepsilon(U), \quad (3.9\text{d})$$

where $\varepsilon(s)$ is the Heaviside step function, the functions $\tilde{f}_{m,\kappa,\omega}(X)$ and $\tilde{\tilde{f}}_{m,\kappa,\omega}(X)$ are determined by

$$\tilde{f}_{m,\kappa,\omega}(X) = \frac{F_{m,\kappa,\omega}}{2} e^{-\pi\omega/(2a)} e^{im\Theta} J_m(\kappa R) \int_{-\infty}^{\infty} \exp\left(iK_Z Z - iWT - \frac{i\omega s}{a}\right) ds, \quad (3.10\text{a})$$

$$\tilde{\tilde{f}}_{m,\kappa,\omega}(X) = \frac{F_{m,\kappa,\omega}}{2} e^{-\pi\omega/(2a)} e^{im\Theta} J_m(\kappa R) \int_{-\infty}^{\infty} \exp\left(iK_Z Z - iWT + \frac{i\omega s}{a}\right) ds, \quad (3.10\text{b})$$

with $K_Z = \kappa \sinh s$ and $W = \kappa \cosh s > 0$. Substituting Eqs. (3.10) into (3.9), one finds

$$\Phi^{(0)}_{\mu;m,\kappa,\omega} = \frac{1}{[1-\exp(-2\pi\omega/a)]^{1/2}} \left(\frac{\partial \tilde{f}_{m,\kappa,\omega}(X)}{\partial T}, 0, 0, \frac{\partial \tilde{f}_{m,\kappa,\omega}(X)}{\partial Z}\right), \quad (3.11\text{a})$$

$$\Phi^{(1)}_{\mu;m,\kappa,\omega} = \frac{1}{[1-\exp(-2\pi\omega/a)]^{1/2}} \left(\frac{\partial \tilde{f}_{m,\kappa,\omega}(X)}{\partial Z}, 0, 0, \frac{\partial \tilde{f}_{m,\kappa,\omega}(X)}{\partial T}\right), \quad (3.11\text{b})$$

$$\Phi^{(2)}_{\mu;m,\kappa,\omega} = \frac{1}{[1-\exp(-2\pi\omega/a)]^{1/2}} \left(0, -\frac{1}{R}\frac{\partial \tilde{f}_{m,\kappa,\omega}(X)}{\partial \Theta}, R\frac{\partial \tilde{f}_{m,\kappa,\omega}(X)}{\partial R}, 0\right), (3.11\text{c})$$

$$\Phi^{(3)}_{\mu;m,\kappa,\omega} = \frac{1}{[1-\exp(-2\pi\omega/a)]^{1/2}} \left(0, \frac{\partial \tilde{f}_{m,\kappa,\omega}(X)}{\partial R}, \frac{\partial \tilde{f}_{m,\kappa,\omega}(X)}{\partial \Theta}, 0\right), \quad (3.11\text{d})$$

The modes $\bar{\Phi}^{(\sigma)}_{\mu;m,\kappa,\omega}$ can be obtained by replacing $\tilde{f}_{m,\kappa,\omega}(X)$ with $\tilde{\tilde{f}}_{m,\kappa,\omega}(X)$ in Eqs. (3.11). We find that modes $\Phi^{(\sigma)}_{\mu;m,\kappa,\omega}$ and $\bar{\Phi}^{(\sigma)}_{\mu;m,\kappa,\omega}$ are superpositions of purely positive-frequency modes in Minkowski space-time, and satisfy the orthonormality and normalization conditions:

$$\left(\Phi^{(\sigma')}_{m',\kappa',\omega'}, \Phi^{(\sigma)}_{m,\kappa,\omega}\right) = -\left(\Phi^{(\sigma')*}_{m',\kappa',\omega'}, \Phi^{(\sigma)*}_{m,\kappa,\omega}\right) = \frac{\eta_{\sigma'\sigma}}{\kappa} \delta_{mm'} \delta(\kappa-\kappa') \delta(\omega-\omega'), \quad (3.12\text{a})$$

$$\left(\Phi^{(\sigma')*}_{m',\kappa',\omega'}, \Phi^{(\sigma)}_{m,\kappa,\omega}\right) = \left(\Phi^{(\sigma')}_{m',\kappa',\omega'}, \Phi^{(\sigma)*}_{m,\kappa,\omega}\right) = 0. \quad (3.12\text{b})$$

Thus, they form a complete basis in Minkowski space-time.

The electromagnetic field $\mathcal{A}_\mu$ can be expanded in Minkowski basis,

$$\mathcal{A}_\mu = \sum_{\sigma=0}^{3}\sum_{m=-\infty}^{\infty}\int_{-\infty}^{\infty}d\omega d\kappa \kappa \left[ b_{m,\kappa,\omega}^{(\sigma)}\Phi_{\mu;m,\kappa,\omega}^{(\sigma)} + b_{m,\kappa,\omega}^{(\sigma)\dagger}\Phi_{\mu;m,\kappa,\omega}^{(\sigma)*} \right], \tag{3.13}$$

the annihilation and creation operators, $b_{m,\kappa,\omega}^{(\sigma)}$ and $b_{m,\kappa,\omega}^{(\sigma)\dagger}$, obey the commutation relations, $\left[ b_{m',\kappa',\omega'}^{(\sigma')}, b_{m,\kappa,\omega}^{(\sigma)\dagger} \right] = (\eta_{\sigma'\sigma}/\kappa)\delta_{mm'}\delta(\kappa-\kappa')\delta(\omega-\omega')$ and $\left[ b_{m',\kappa',\omega'}^{(\sigma')}, b_{m,\kappa,\omega}^{(\sigma)} \right] = \left[ b_{m',\kappa',\omega'}^{(\sigma')\dagger}, b_{m,\kappa,\omega}^{(\sigma)\dagger} \right] = 0$. Then, the relationship of the annihilation and creation operators between the Rindler modes and Minkowski modes are obtained

$$a_{m,\kappa,\omega}^{R(\sigma)} = \frac{b_{m,\kappa,\omega}^{(\sigma)} + \exp\left(-\frac{\pi\omega}{a}+im\pi\right)b_{-m,\kappa,-\omega}^{(\sigma)\dagger}}{\left[1-\exp\left(-\frac{2\pi\omega}{a}\right)\right]^{1/2}}, \quad a_{m,\kappa,\omega}^{R(\sigma)\dagger} = \frac{b_{m,\kappa,\omega}^{(\sigma)\dagger} + \exp\left(-\frac{\pi\omega}{a}-im\pi\right)b_{-m,\kappa,-\omega}^{(\sigma)}}{\left[1-\exp\left(-\frac{2\pi\omega}{a}\right)\right]^{1/2}},$$

(3.14a)

$$a_{m,\kappa,\omega}^{L(\sigma)} = \frac{b_{m,\kappa,-\omega}^{(\sigma)} + \exp\left(-\frac{\pi\omega}{a}+im\pi\right)b_{-m,\kappa,\omega}^{(\sigma)\dagger}}{\left[1-\exp\left(-\frac{2\pi\omega}{a}\right)\right]^{1/2}}, \quad a_{m,\kappa,\omega}^{L(\sigma)\dagger} = \frac{b_{m,\kappa,-\omega}^{(\sigma)\dagger} + \exp\left(-\frac{\pi\omega}{a}-im\pi\right)b_{-m,\kappa,\omega}^{(\sigma)}}{\left[1-\exp\left(-\frac{2\pi\omega}{a}\right)\right]^{1/2}}$$

(3.14b)

Based on Eq. (3.14), the expectation value of the Rindler particle number operator, $\hat{N}_{m,\kappa,\omega}^{R(\sigma)} = \zeta_\sigma a_{m,\kappa,\omega}^{R(\sigma)\dagger}a_{m,\kappa,\omega}^{R(\sigma)}$, in Minkowski vacuum is derived,

$$\left\langle 0_M \left| \hat{N}_{m,\kappa,\omega}^{R(\sigma)} \right| 0_M \right\rangle = \frac{1}{\exp(2\pi\omega/a)-1}. \tag{3.15}$$

It is found from Eq. (3.15) that the number expectation value only depends on the photon energy, and it is a Bose-Einstein distribution with the temperature $T=a/(2\pi)$. That means when an observer moves with a uniform proper acceleration in Minkowski space-time, it experiences the vacuum as a thermal bath in equilibrium state. This conclusion is analogous to the case of scalar field.

### 3.3 Physical effects in MF
### 3.3.1 Absorption and emission rates from a vortex current in thermal bath

Next, we discuss the possible observable effect of the photon Unruh effect related to the OAM, which is determined by the interaction of a vortex electric current with the background thermal bath. Let us consider a current generated by an electron beam travelling along a hyperbolic world line with a uniform proper acceleration $a$. This electron beam has a vortex structure in the transverse dimensions with a spiral azimuthal phase, $\cos(l\theta - \Omega t)$, where $l$ is the OAM quantum number, $\Omega > 0$ determines the angular velocity of the vortex current. Under these assumptions, the four-current density can be expressed as $j^\mu(x) = (q/2)(r/w)^{|l|} \exp(-r^2/w^2) \delta(z)(1 + \cos(l\theta - \Omega t))(1, 0, \Omega/l, 0)$, where $q$ is the charge surface density, $w$ is the waist radius of the current. This current satisfies the conservation law: $\nabla_\mu j^\mu(x) = 0$. The static part of the current (the first term) only produces the absorption and emission of zero-energy photons [15] without OAM, and we will not discuss it. What we are interested in is the vortex part,

$$j^\mu_{vor}(x) = \frac{q}{2}\left(\frac{r}{w}\right)^{|l|} \exp\left(-\frac{r^2}{w^2}\right) \delta(z) \cos(l\theta - \Omega t)\left(1, 0, \frac{\Omega}{l}, 0\right). \quad (3.16)$$

The interaction Hamiltonian of the current with the background field in the thermal bath is expressed as $H_{int} = \int j^\mu(x) \mathcal{A}_\mu(x) \sqrt{-g} d^3x$. The S-operator of the interaction takes the form, $\hat{S}(t,t_0) = T\exp\left[-i \int_{t_0}^{t} j^\mu(x') \mathcal{A}_\mu(x') \sqrt{-g(x')} d^3x' dt'\right]$. In the first order approximation, the amplitude for absorbing a Rindler photon with mode $|\sigma, m, \kappa, \omega\rangle_R$ from the thermal bath by the current is given by $S^R_{absorb}(\sigma, m, \kappa, \omega) = -i\zeta^{N^{(\sigma)}_{m,n,\omega}}_\sigma \sqrt{N^{(\sigma)}_{m,n,\omega}} \int d^4x \sqrt{-g(x)} j^\mu_{vor}(x) A^{R(\sigma)}_{\mu;m,n,\omega}(x)$. For a realistic physical process, only the physical modes, $\sigma = 1, 2$, should be considered. Using Eqs. (3.2) and (3.16), one obtains the amplitude for absorbing a Rindler photon with the physical mode ($\sigma$=1) from the thermal bath,

$$S^R_{absorb}(1, m, \kappa, \omega) = \frac{i\pi^2 q F_{m,\kappa,\omega} \kappa^{|l|+1} w^{|l|+2}}{2^{|l|+2}} \sqrt{N^{(1)}_{m,\kappa,\omega}} \exp\left(-\frac{\kappa^2 w^2}{4}\right) \\ \times \left[K_{i\omega/a-1}\left(\frac{\kappa}{a}\right) + K_{i\omega/a+1}\left(\frac{\kappa}{a}\right)\right] \delta_{ml} \delta(\omega - \Omega) \quad (3.17a)$$

and the amplitude for absorbing a Rindler photon with the physical mode ($\sigma$=2),

$$S_{absorb}^{R}(2,m,\kappa,\omega) = -\frac{i\pi^2 q\Omega F_{m,\kappa,\omega} \kappa^{|l|} w^{|l|+2}}{2^{|l|+1} l} \sqrt{N_{m,\kappa,\omega}^{(2)}} \left(|l| - \frac{\kappa^2 w^2}{2}\right)$$
$$\times \exp\left(-\frac{\kappa^2 w^2}{4}\right) K_{i\omega/a}\left(\frac{\kappa}{a}\right) \delta_{ml}\delta(\omega - \Omega) \quad . \quad (3.17b)$$

The differential absorption rate per unit time can be calculated by using

$$P_{absorb}^{R} = \frac{1}{T}\sum_{m}\int d\omega d\kappa \kappa \sum_{\sigma=1}^{2} \left|S_{absorb}^{R}(\sigma,m,\kappa,\omega)\right|^2,$$ with the result

$$\frac{dP_{absorb}^{R}}{d\kappa} = \frac{q^2 w^{2|l|+4}}{2^{2|l|+5} a} \exp\left(-\frac{\pi\Omega}{a}\right) \kappa^{2|l|-1} \exp\left(-\frac{\kappa^2 w^2}{2}\right)$$
$$\times \left\{\frac{\kappa^2}{4}\left[K_{i\Omega/a-1}\left(\frac{\kappa}{a}\right) + K_{i\Omega/a+1}\left(\frac{\kappa}{a}\right)\right]^2 + \frac{\Omega^2}{l^2}\left(|l| - \frac{\kappa^2 w^2}{2}\right)^2 K_{i\Omega/a}^2\left(\frac{\kappa}{a}\right)\right\} \quad . \quad (3.18)$$

The amplitude for emitting a Rindler photon with mode $|\sigma,m,\kappa,\omega\rangle_R$ to the thermal bath by the vertex current is given by $S_{emit}^{R}(\sigma,m,\kappa,\omega) = -i\zeta_{\sigma}^{N_{m,\kappa,\omega}^{(\sigma)}+1}\sqrt{N_{m,\kappa,\omega}^{(\sigma)}+1}\int d^4x\sqrt{-g(x)}j_{vor}^{\mu}(x)A_{\mu;m,\kappa,\omega}^{R(\sigma)*}(x)$. Based on Eqs. (3.2) and (3.16) one derives the emission amplitude for the physical mode ($\sigma=1$),

$$S_{emit}^{R}(1,m,\kappa,\omega) = \frac{i\pi^2 q F_{m,\kappa,\omega} \kappa^{|l|+1} w^{|l|+2}}{2^{|l|+2}} \sqrt{N_{m,\kappa,\omega}^{(1)}+1} \exp\left(-\frac{\kappa^2 w^2}{4}\right)$$
$$\times \left[K_{i\omega/a-1}\left(\frac{\kappa}{a}\right) + K_{i\omega/a+1}\left(\frac{\kappa}{a}\right)\right]\left(\frac{l}{|l|}\right)^{|l|} \delta_{ml}\delta(\omega-\Omega) \quad , \quad (3.19a)$$

and for the physical mode ($\sigma=2$),

$$S_{emit}^{R}(2,m,\kappa,\omega) = -\frac{i\pi^2 q\Omega F_{m,\kappa,\omega} \kappa^{|l|} w^{|l|+2}}{2^{|l|+1} l} \sqrt{N_{m,\kappa,\omega}^{(2)}+1}\left(|l| - \frac{\kappa^2 w^2}{2}\right)$$
$$\times \exp\left(-\frac{\kappa^2 w^2}{4}\right) K_{i\omega/a}\left(\frac{\kappa}{a}\right)\left(\frac{l}{|l|}\right)^{|l|} \delta_{ml}\delta(\omega-\Omega) \quad . \quad (3.19b)$$

Then the emission rate per unit time is obtained,

$$\frac{dP_{emit}^{R}}{d\kappa} = \frac{q^2 w^{2|l|+4}}{2^{2|l|+5} a} \exp\left(\frac{\pi\Omega}{a}\right) \kappa^{2|l|-1} \exp\left(-\frac{\kappa^2 w^2}{2}\right)$$
$$\times \left\{\frac{\kappa^2}{4}\left[K_{i\Omega/a-1}\left(\frac{\kappa}{a}\right) + K_{i\Omega/a+1}\left(\frac{\kappa}{a}\right)\right]^2 + \frac{\Omega^2}{l^2}\left(|l| - \frac{\kappa^2 w^2}{2}\right)^2 K_{i\Omega/a}^2\left(\frac{\kappa}{a}\right)\right\} \quad . \quad (3.20)$$

From Eqs. (3.17) and (3.19) we conclude that if the current density has a vertex structure and carries a definite OAM of $l$, it can absorb and emit Rindler photons with the energy $\omega = \Omega$ and OAM of $l$ when it interacts with the background thermal bath. The absorption rate is equal to the emission rate only when the rotational velocity $\Omega \to 0$.

**3.3.2 Photon emission rate from the current in Minkowski vacuum**

In the Minkowski space-time, the current density can be obtained according to the vectorial transformation rules: $J_{vor}^{\mu}(X) = (\partial X^{\mu}/\partial x^{\nu}) j_{vor}^{\mu}(\Lambda^{-1}x)$, the vortex part reads

$$J_{vor}^{\mu}(X) = \frac{q}{2}\left(\frac{R}{w}\right)^{|l|} \exp\left(-\frac{R^2}{w^2}\right) \delta(z) \cos(l\Theta - \Omega t)(aZ,\ 0,\ \Omega/l,\ aT). \quad (3.21)$$

The rate of emitting a photon from the current is calculated as $P_{emit}^{M} = \frac{1}{T}\int d^3k \sum_{\sigma}\left|S_{emit}^{M}(\sigma,\mathbf{k},\omega)\right|^2$, where $S_{emit}^{M}(\sigma,\mathbf{k},\omega)$ is the emission amplitude for the mode $(\sigma,\mathbf{k},\omega)$ and given by $S_{emit}^{M}(\sigma,\mathbf{k},\omega) = \frac{-i}{\sqrt{2\omega}}\int d^4X J_{vor}^{\mu}(X) \varepsilon_{\mu}^{(\sigma)*} e^{i\omega T - i\mathbf{k}\cdot\mathbf{X}}$, $\varepsilon_{\mu}^{(\sigma)}$ is the polarization vector of the emitted photon. By using the polarization sum formula of photon, $\sum_{\sigma}\varepsilon_{\mu}^{(\sigma)*}\varepsilon_{\nu}^{(\sigma)} = g_{\mu\nu}$, one obtains the emission rate,

$$P_{emit}^{M} = \frac{1}{T}\int \frac{d^3k}{(2\pi)^3 2\omega}\int d^4X d^4X' J_{vor}^{\mu}(X) J_{vor,\mu}^{*}(X') e^{i\mathbf{k}\cdot(\mathbf{X}'-\mathbf{X})-i\omega(T'-T)},$$ which leads to the differential emission rate,

$$\frac{dP_{emit}^{M}}{d\kappa} = \frac{q^2 w^{2|l|+4}}{2^{2|l|+3} a}\cosh\left(\frac{\pi\Omega}{a}\right)\kappa^{2|l|-1}\exp\left(-\frac{\kappa^2 w^2}{2}\right)$$
$$\times \left[\frac{\kappa^2}{4}\left(K_{i\Omega/a+1}^{2}\left(\frac{\kappa}{a}\right) + K_{i\Omega/a-1}^{2}\left(\frac{\kappa}{a}\right)\right) + \frac{\Omega^2}{l^2}\left(\left(|l| - \frac{\kappa^2 w^2}{4}\right)^2 + \frac{\kappa^4 w^4}{16}\right)K_{i\Omega/a}^{2}\left(\frac{\kappa}{a}\right)\right].$$

$$(3.22)$$

Using the recurrence relations of the modified Bessel function, $(2\nu/s)K_{\nu}(s) = K_{\nu+1}(s) - K_{\nu-1}(s)$, we find the relation: $dP_{emit}^{M}/d\kappa = dP_{absorb}^{R}/d\kappa + dP_{emit}^{R}/d\kappa$, indicating that the photon emission rate from an accelerated vortex current carrying definite OAM of $l$ in Minkowski vacuum can be reproduced by summing the absorption and emission rates of Rindler photons with energy of $\omega = \Omega$ and OAM of $m = l$ in Unruh thermal bath.

### 3.3.3 The definition of twisted photon state

Although the calculation in Section 3.3.2 gives the expected emission rate in MF, it does not reveal the details of the OAM characteristics. To reveal the OAM correlation between the emitted photons in FM and the absorbed and emitted Rindler photons, we need to define the photon state with definite OAM, i.e., the twisted photon state, in MF. By using three polarization base vectors [21]: $\xi^{(\sigma)\mu} = -\frac{\sigma}{\sqrt{2}}(0,1,i\sigma,0)$ and $\xi^{(z)\mu} = (0,0,0,1)$ with helicity $\sigma = \pm 1$, a twisted photon state can be defined as $\mathcal{A}_{j,\kappa,\omega}^{(\sigma)\mu} = \sqrt{c}/\left(4\pi^2\kappa\sqrt{2\omega}\right) e^{ik_Z Z - i\omega T} \int \varepsilon_{k_\perp}^{(\sigma)\mu}(-i)^j e^{ij\varphi} \delta(k_\perp - \kappa) e^{i\mathbf{k}_\perp \cdot \mathbf{R}} d\varphi k_\perp dk_\perp$, where the polarization vector $\varepsilon_{k_\perp}^{(\sigma)\mu} = \xi^{(-\sigma)\mu} e^{i\sigma\varphi} \sin^2(\vartheta/2) + \xi^{(\sigma)\mu} e^{-i\sigma\varphi} \cos^2(\vartheta/2) + \left(\sigma/\sqrt{2}\right)\xi^{(z)\mu}\sin\vartheta$ satisfies $k_\mu \varepsilon_{k_\perp}^{(\sigma)\mu} = 0$ and $\varepsilon_{k_\perp}^{(\sigma)\mu}\varepsilon_{\mu,k_\perp}^{(\sigma')*} = \delta_{\sigma\sigma'}$, $k_\perp$ is the photon transverse momentum, $\vartheta$ and $\varphi$ are the polar and azimuth angles of the photon momentum and $\tan\vartheta = \kappa/k_Z$. In cylindrical coordinates the twisted photon state takes the from,

$$\mathcal{A}_{\mu;j,\kappa,\omega}^{(\sigma)} = \frac{e^{ij\Theta + ik_Z Z - i\omega T}}{4\pi\sqrt{\omega}} \begin{pmatrix} 0 \\ i\sin^2(\vartheta/2) J_{j+\sigma}(\kappa R) + i\cos^2(\vartheta/2) J_{j-\sigma}(\kappa R) \\ \sigma R \sin^2(\vartheta/2) J_{j+\sigma}(\kappa R) - \sigma R \cos^2(\vartheta/2) J_{j-\sigma}(\kappa R) \\ \sigma \sin\vartheta J_j(\kappa R) \end{pmatrix}.$$

(3.23)

It is not difficult to prove that the twisted photon state satisfies the orthonormality: $\left(\mathcal{A}_{j',\kappa',\omega'}^{(\sigma')}, \mathcal{A}_{j,\kappa,\omega}^{(\sigma)}\right) = \frac{1}{\kappa}\delta_{\sigma\sigma'}\delta_{jj'}\delta(\kappa-\kappa')\delta(k_Z - k_Z')$. It is the eigen mode of total angular momentum (TAM) operator in Z-direction with the eigen value $j$, but not the OAM and SAM operators.

### 3.3.4 Recalculation of the emission rate of twisted photons

Now, we can recalculate the emission rate of the twisted photons from the accelerated vortex current in Minkowski vacuum. In the first order approximation, the amplitude for emitting a twisted photon with mode $|\sigma, j, \kappa, k_z, \omega\rangle_M$ from the current is given by $S_{emit}^M(\sigma, j, \kappa, k_z, \omega) = -i\int d^4X R J_{vor}^\mu(X) \mathcal{A}_{\mu;j,\kappa,k_z,\omega}^{(\sigma)*}(X)$. Using the current density (3.21) and twisted modes (3.23), the emission amplitude of the twisted photon is obtained,

$$S_{emit}^M(\sigma, j, \kappa, k_z, \omega)$$
$$= -\frac{iq\kappa^{|l|-1}w^{|l|+2}}{2^{|l|+4}a\sqrt{\omega}}\exp\left(-\frac{\kappa^2 w^2}{4}\right)$$
$$\times \left\{\left(\frac{l}{|l|}\right)^{|l|}\delta_{jl}\exp\left(\frac{\pi\Omega}{2a}\right)\left(\frac{\omega+k_z}{\omega-k_z}\right)^{-i\Omega/(2a)}\left[\frac{2\Omega}{l}\left(\frac{\kappa^2 w^2}{2}-\left(|l|+\frac{\sigma l k_z}{\omega}\right)\right)K_{i\Omega/a}\left(\frac{\kappa}{a}\right)\right.\right.$$
$$\left.+\frac{i\sigma\kappa^2}{\omega}\left(\left(\frac{\omega+k_z}{\omega-k_z}\right)^{1/2}K_{i\Omega/a-1}\left(\frac{\kappa}{a}\right)+\left(\frac{\omega+k_z}{\omega-k_z}\right)^{-1/2}K_{i\Omega/a+1}\left(\frac{\kappa}{a}\right)\right)\right]$$
$$+\left(-\frac{l}{|l|}\right)^{|l|}\delta_{j,-l}\exp\left(-\frac{\pi\Omega}{2a}\right)\left(\frac{\omega+k_z}{\omega-k_z}\right)^{i\Omega/(2a)}\left[\frac{2\Omega}{l}\left(\frac{\kappa^2 w^2}{2}-\left(|l|-\frac{\sigma l k_z}{\omega}\right)\right)K_{i\Omega/a}\left(\frac{\kappa}{a}\right)\right.$$
$$\left.\left.+\frac{i\sigma\kappa^2}{\omega}\left(\left(\frac{\omega+k_z}{\omega-k_z}\right)^{1/2}K_{i\Omega/a+1}\left(\frac{\kappa}{a}\right)+\left(\frac{\omega+k_z}{\omega-k_z}\right)^{-1/2}K_{i\Omega/a-1}\left(\frac{\kappa}{a}\right)\right)\right]\right\}$$

(3.24)

Eq. (3.24) indicates that the emitted photons carry the well-defined OAM of $j = \pm l$. Then, the differential rate for emitting a twisted photon per unit time can be calculated using $P_{emit}^M = \frac{1}{\mathcal{T}}\sum_j \int dk_z d\kappa\kappa \sum_{\sigma=\pm 1}|S_{emit}^M(\sigma, j, \kappa, k_z, \omega)|^2$, with the result

$$\frac{dP_{emit}^M}{d\kappa} = \frac{q^2 w^{2|l|+4}}{2^{2|l|+4}a}\cosh\left(\frac{\pi\Omega}{a}\right)\kappa^{2|l|-1}\exp\left(-\frac{\kappa^2 w^2}{2}\right)$$
$$\times \left\{\frac{\kappa^2}{4}\left[K_{i\Omega/a-1}\left(\frac{\kappa}{a}\right)+K_{i\Omega/a+1}\left(\frac{\kappa}{a}\right)\right]^2 + \frac{\Omega^2}{l^2}\left(|l|-\frac{\kappa^2 w^2}{2}\right)^2 K_{i\Omega/a}^2\left(\frac{\kappa}{a}\right)\right\}. \quad (3.25)$$

Now we reproduce the results, $P_{emit}^M = P_{absorb}^R + P_{emit}^R$, by using the twisted photon state, based on which we come to a conclusion, an accelerated vortex current carrying definite OAM of $l$ can emit twisted photons with TAM of $j = \pm l$ in Minkowski vacuum, the corresponding emission rate can be reproduced by summing the absorption and emission rates of Rindler photons with the energy of $\omega=\Omega$ and OAM of $m=l$. The emission rate of a Rindler photon with the OAM of $l$ is equal to the emission rate of a twisted photon with the TAM of $l$ in Minkowski vacuum, $P_{emit}^{M}(j=l)=P_{emit}^{R}(m=l)$, but the absorption rate of a Rindler photon is equal to the emission rate of a twisted photon with the TAM of $-l$ in Minkowski vacuum, $P_{emit}^{M}(j=-l)=P_{absorb}^{R}(m=l)$, which can be found in Eq. (3.24).

Next, we give a short discussion on numerical results. Suppose the surface density of an electron current is $n_0=4.5\times10^{18}$cm$^{-2}$, leading to the charge surface density $q=en_0$, the waist radius $w=3\times10^{-4}$cm. Fig. 1(a) shows the differential emission rate of twisted photons in MF with different OAM number of vortex current, $l=1$, 2 and 3, where the proper acceleration of the vortex current is $a=3\times10^{22}$cm/s$^2$, and the proper angular velocity $\Omega=3\times10^{13}$rad/s. We find that the emission rate decreases rapidly with the increasing of $l$, because the increased OAM number will decrease the angular velocity of the vortex current when $\Omega$ is constant, see the spiral phase in the current, $\cos(l\theta-\Omega t)$. If we increase $\Omega$, the emission rate will be increased remarkably, as shown in Fig. 1(b) with $l=1$. However, the higher angular velocity leads to the faster linear velocity of the rotational electron in vortex current, which is limited by the light speed. We use $2w$ to calculate the radius of the rotating trajectory of the outermost electron, its rotational velocity is $v_{out}=2w\Omega$. For the waist radius $w=3\times10^{-4}$cm, the maximum $\Omega$ is $5\times10^{13}$ rad/s. Fig. 1(c) shows the emission rate with different proper acceleration, $a=3\times10^{22}$cm/s$^2$, $10^{23}$cm/s$^2$, and $5\times10^{23}$cm/s$^2$, the OAM number is $l=1$ and angular velocity $\Omega=3\times10^{13}$ rad/s here. It is found that the higher acceleration can broaden the spectrum and eliminate the oscillation.

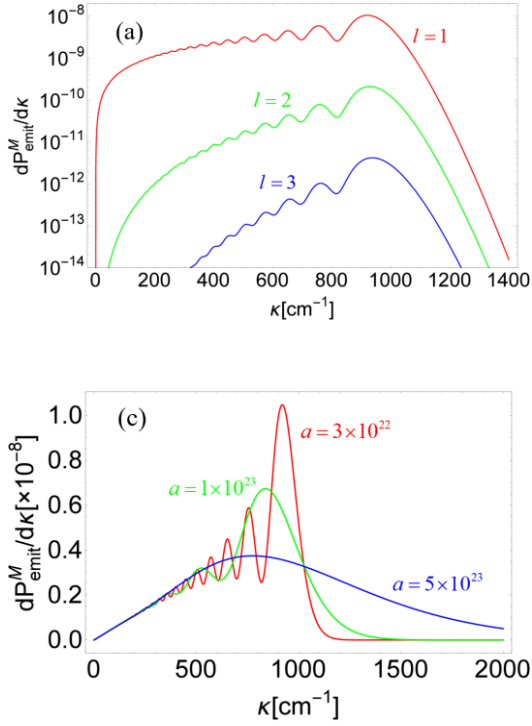

Fig.1. (a) Emission rate of twisted photons when the vortex current possessing OAM of $l=1,2,3$, proper acceleration $a=3\times10^{22}$cm/s$^2$, and proper angular velocity $\Omega=3\times10^{13}$rad/s. (b) Emission rate of twisted photons with proper angular velocity of the current $\Omega=10^{13}$, $3\times10^{13}$, $4.5\times10^{13}$rad/s, and OAM $l=1$. (c) Emission rate of twisted photons with proper angular velocity of the current $\Omega=3\times10^{13}$rad/s, and proper acceleration $a=3\times10^{22}$, $10^{23}$, $5\times10^{23}$cm/s$^2$.

## IV Conclusion

In conclusion, we have demonstrated theoretically that the rotation and angular momentum can induce the observably physical phenomena associated with the Unruh effect, although they do not affect the background temperature of the thermal bath which is only determined by the longitudinal proper acceleration. If an observer has a rotational vortex structure in the transverse dimensions and carries the well-defined OAM, it can absorb or emit Rindler particles with the same OAM when interacting with the background thermal bath. The energies of the absorbed and emitted particles are determined by the angular velocity of the vortex observer. In general, the emission rate is not identical to the absorption rate. Only when the proper angular velocity $\Omega\to 0$, we get the identical absorption and emission rates, which is consistent with the static observer interacting with the zero-energy Rindler particles in the thermal bath [15]. In Minkowski vacuum, the emission process of Rindler particles is connected to the emission of particles carrying the same angular momentum with the observer; but the absorption process of the Rindler particles is connected to the emission of particles carrying the opposite angular momentum. The emission rate in Minkowski vacuum can be reproduced by summing the absorption and emission rates of Rindler particles. The electromagnetic field is a vector field, so photon carries both OAM and SAM. If we use an accelerated vortex current as the observer, the absorbed/emitted Rindler photons are connected to the emitted photons with the TAM opposite/identical to the OAM of vortex current. The discussed physical results establish the direct correlation between the observable phenomena in laboratory reference frame and Unruh effect. Especially, the angular momentum provides a new and effective degree of freedom to experimentally filter out background noise so that the signal could be extracted sufficiently, so it has a potential prospect for the detection of Unruh effect.

**Acknowledgments**

This work is supported by the Ministry of Science and Technology of the People's Republic of China (Grant Nos. 2018YFA0404803 and 2016YFA0401102), the National Natural Science Foundation of China (Nos. 11875307, 11935008 and 11774415), the Strategic Priority Research Program of Chinese Academy of Sciences (Grant No. XDB16010000).